\documentclass[12pt]{article}
\usepackage[margin=1in]{geometry}
\usepackage{hyperref}
\usepackage{bbding}

\usepackage[english]{babel}
\usepackage{amsmath,amsthm}
\usepackage{amssymb}
\usepackage{amsfonts}

\usepackage{algorithm}  
\usepackage{algorithmicx}  
\usepackage{algpseudocode}  

\usepackage{color}

\usepackage{graphicx}
\usepackage{pgf,tikz}
\usetikzlibrary{shapes,arrows,automata}

\newtheorem{thm}{Theorem}[section]

\newtheorem{cor}[thm]{Corollary}
\newtheorem{lem}[thm]{Lemma}
\newtheorem{conj}[thm]{Conjecture}
\newtheorem{ques}[thm]{Question}
\newtheorem{prop}[thm]{Proposition}
\theoremstyle{definition}
\newtheorem{defn}[thm]{Definition}

\newtheorem{fact}[thm]{Fact}
\numberwithin{equation}{section}

\newcommand{\sfL}{\mathsf{L}}

\begin{document}
\title{Direct Sum  Theorems From Fortification}%
\author{Hao Wu\thanks{College of Information Engineering, Shanghai Maritime University, Shanghai, China. My email is \texttt{haowu@shmtu.edu.cn}, you can also reach me via \texttt{wealk@outlook.com}.}}%
\maketitle
\begin{abstract}We revisit  the direct sum questions in communication complexity which asks whether the resource needed to solve $n$ communication problems together is (approximately) the sum of resources needed to solve these problems separately. Our work starts with the  observation that Dinur and Meir's fortification lemma \cite{DBLP:journals/cc/DinurM18} can be generalized to a general fortification lemma for a sub-additive measure over set. By applying this lemma to the case of cover number, we obtain a dual form of cover number, called ``$\delta$-fooling set'' which is a generalized fooling set. Any rectangle which contains enough number of elements from a $\delta$-fooling set  can not be monochromatic.
	
With this fact, we are able to reprove the classic direct sum theorem \cite{DBLP:journals/siamdm/KarchmerKN95} of cover number with a simple double counting argument. Formally, let  $S  \subseteq (A\times B) \times O$ and $T \subseteq (P\times Q) \times Z$ be two communication problems, $
\log \mathsf{Cov}\left(S\times T\right)
\geq  \log \mathsf{Cov}\left(S\right) 
+ \log\mathsf{Cov}(T)
-\log\log|P||Q|-4.$ where $\mathsf{Cov}$ denotes the cover number. One issue of current deterministic direct sum theorems \cite{DBLP:journals/siamcomp/FederKNN95,DBLP:journals/siamdm/KarchmerKN95} about communication complexity is that they provide no information when $n$ is small, especially when $n=2$. In this work,  we prove a new direct sum theorem about protocol size which imply a better direct sum theorem for two functions in terms of protocol size. Formally, let $\mathsf{L}$ denotes complexity of  the protocol size of  a communication problem, given a communication problem $F:A \times B \rightarrow 
 \{0,1\}$, $
 \log\mathsf{L}\left(F\times F\right)\geq
 \log \mathsf{L}\left(F\right) +\Omega\left(\sqrt{\log\mathsf{L}\left(F\right)}\right)-\log\log|A||B| -4$. All our  results are obtained in a similar way using the $\delta$-fooling set to construct a hardcore for the direct sum problem.

\end{abstract}

\newpage
\tableofcontents
\newpage

\section{Introduction}
The  direct sum question, in general, asks following question: whether the resource to solve several tasks together is (approximately) the sum of resources to solve these tasks separately. Particularity, we want to know whether the resource to solve $n$ copies of certain task is (approximately) $n$ times  the resource to solve one copy of the  task. In the field of communication complexity, the direct sum question was proposed by  Karchmer, Raz and Wigderson \cite{DBLP:journals/cc/KarchmerRW95}. Let communication problem $F:X\times Y \rightarrow \{0,1\}$ be a function and $\mathsf{C}$ denote the communication complexity of a problem, current known deterministic direct sum theorems\cite{DBLP:journals/siamcomp/FederKNN95,DBLP:journals/siamdm/KarchmerKN95}\footnote{See \cite{Pankratov-master} for a more detailed introduction of direct sum problems in communication complexity.} tell us solving $n$ copies of communication  problem $F$ requires $\Omega\left(n\cdot \sqrt{\mathsf{C}(F)} \right)$ bits of communication. But in some cases, current direct sum theorems are not satisfactory. When $n$ is small, especially when $n=2$,  current direct sum theorems tell us nothing more than the trivial lower bound. Intuitively, we should be able to prove  stronger results when $n=2$, that if the communication complexity of $F$ is not too small, then solving two copies of $F$ should require strictly more bits of communication than solving only one copy of $F$. Formally, we have following conjecture which is consistent with current known direct sum theorems.
\begin{conj}Let communication problem $F:X\times Y \rightarrow \{0,1\}$ be a function, $\mathsf{C}(F\times F) \geq \mathsf{C}(F) + \Omega\left(\sqrt{\mathsf{C}(F)} \right)$.
\end{conj}
Let $\mathsf{L}(F)$ denote the complexity of the protocol size of the communication problem $F:X\times Y \rightarrow \{0,1\}$, similarly, we have  following conjecture about protocol size.
\begin{conj}Let communication problem $F:X\times Y \rightarrow \{0,1\}$ be a function, $\log\mathsf{L}(F\times F) \geq \log\mathsf{L}(F) + \Omega\left(\sqrt{\log\mathsf{L}(F)} \right)$.
\end{conj}
Finally, in all  direct sum question, particularly, in the case of cover number, an ideal situation is that there is a `hardcore' of cover number,  when we know some information about this hardcore, the residual hardness is just the result of the  complexity  of the hardcore minus the amount of  known information.  A potential hardcore of cover number may be some kind of generalized form of standard fooling set, it is well known that large standard fooling set implies large cover number, but not vice versa. So naturally, we have following question.
\begin{ques}Is there a generalized form of fooling set which can be viewed as a dual form of cover number? Particularly, is large cover number implies some kind of fooling set?
\end{ques}

In this paper, we tackle these questions and make some progress about them.
\subsection{Our results}
All our results rely on following fact: there is a  dual form of cover number, $\delta$-fooling set, a concept generalized from the standard fooling set. Formally,  we have following concept. 
\begin{defn}[$\delta$-fooling set]Let  $S\subseteq (X\times Y) \times Z$ be a communication problem, we call a set  $\Lambda \subseteq  X\times Y$ a $\delta$-fooling set of $S$ if for any subset $\tilde{\Lambda} \subseteq \Lambda$ such that $\frac{|\tilde{\Lambda}|}{|\Lambda|} > \delta$, there is no  monochromatic rectangle that covers all elements in the subset $\tilde{\Lambda}$.
\end{defn}	
Note  that a standard fooling set $\Lambda$ is just a special case of $\delta$-fooling set where $\delta= 1/|\Lambda|$.
It is easy to see that a $\delta$-fooling set with small $\delta$ implies a large cover number lower bound. The harder direction is to show  a large cover number implies  a $\delta$-fooling set with small $\delta$. This is done by the technique of ``fortification''. The term ``fortification'' is introduced by Moshkovitz \cite{DBLP:conf/focs/Moshkovitz14}  to prove a parallel repetition theorem. Then Dinur and Meir\cite{DBLP:journals/cc/DinurM18} introduced this idea into communication complexity and proved a fortification lemma of protocol size over rectangles. The spirit of this concept is that given some hard problem, we want to `fortify' it into a hardcore such that if we already know some information about this hardcore, the residual hardness is just the result of the  complexity  of the hardcore minus the amount of  known information. We observer that   Dinur and Meir's fortification lemma can be generalized into a general form then apply it to the case of cover number. Formally, we have following result about fortification of cover number thus complete the harder direction of the duality.
\begin{prop}[Fortification of cover number]Let  $S\subseteq (X\times Y) \times Z$ be a communication problem,  there exists $\Lambda \subseteq  X\times Y$ such that for any subset $\tilde{\Lambda} \subseteq \Lambda$, we have
	$$\mathsf{Cov}(\tilde{\Lambda}) \geq
	\frac{|\tilde{\Lambda}|}{|\Lambda|}
	\cdot\frac{ \mathsf{Cov}(S)}{16\log|X| |Y|}. $$
	Particularly, when $	\frac{|\tilde{\Lambda}|}{|\Lambda|}> \frac{16\log|X||Y|}{\mathsf{Cov}(S)}$, we have  $\mathsf{Cov}(\tilde{\Lambda}) > 1$, this means $\Lambda$ is a $\frac{16\log|X| |Y|}{\mathsf{ Cov}(S)}$-fooling set.
\end{prop}
 Use this dual form of cover number, we are able to reprove  the direct sum theorem  of cover number.
 \begin{thm}Given two communication problems $S  \subseteq (A\times B) \times O$ and $T \subseteq (P\times Q) \times Z$, we have
 	$
 	\log \mathsf{Cov}\left(S\times T\right)
 	\geq  \log \mathsf{Cov}\left(S\right) 
 	+ \log\mathsf{Cov}(T)
 	-\log\log|P||Q|-4.
 	$
 \end{thm}
Along the way, we prove a new direct sum result of protocol size which imply a better direct sum theorem for two functions in terms of protocol size. This gives a positive answer to Conjecture 1.2.
\begin{thm}Given two communication problem $S \subseteq (A\times B) \times O$ and $T\subseteq (P\times Q) \times Z$, we have
	$
	\log\mathsf{L}\left(S\times T\right) \geq \log \mathsf{L}\left(S\right) + \log\mathsf{Cov}(T)-\log\log|P||Q| -4
	$. 
\end{thm}
\begin{cor}Given a communication problem $F:A \times B \rightarrow 
	\{0,1\}$, 
\begin{align*}
\log\mathsf{L}\left(F\times F\right)\geq
\log \mathsf{L}\left(F\right) +\Omega\left(\sqrt{\log\mathsf{L}\left(F\right)}\right)-\log\log|A||B| -4.
\end{align*}	
\end{cor}
For conjecture 1.1, our approach does not work due to that the measure of  communication complexity is less  structural than the measure of protocol size, and we leave this conjecture as an interesting open problem.

\subsection{Our approach}
In this section, at first, we show how to generalize  Dinur and Meir's fortification lemma with the right abstraction. Then by applying this general fortification lemma to the case of cover number, we have the fortification lemma of cover number and existence of $\delta$-fooling set.  Secondly, we take the direct sum theorem of cover number as a running example of proving direct sum type theorem using $\delta$-fooling set. The direct sum theorem of protocol size follows a similar paradigm,  that is using the $\delta$-fooling set to construct a hardcore for the direct sum  problem.

  Dinur and Meir's fortification lemma can be generalized into following setting: let $\Sigma$ be a nonempty finite set,  a sub-additive measure over $\Sigma$ is a function  $\mu: 2^\Sigma \rightarrow \mathbb{N}$ with following properties:
\begin{itemize}
	\item semipositivity: 
	$\mu(\emptyset)=0$ and 	if $\Lambda \subseteq \Sigma$  is not empty, $\mu(\Lambda)\geq 1$,
	\item subadditivity: given two subsets $\Lambda,\Lambda^\prime \subseteq \Sigma$, $\mu(\Lambda \cup \Lambda^\prime)\leq\mu(\Lambda)+\mu(\Lambda^\prime).$
\end{itemize}
And the general fortification lemma states following fact: there is a subset $\Lambda \subseteq \Sigma$ such that give any subset $\tilde{\Lambda} \subseteq \Lambda$, it holds that
\[ \mu(\tilde{\Lambda}) \geq \frac{1}{4\log |\Sigma|} \cdot  \frac{|\tilde{\Lambda}|}{|\Lambda|} \cdot \mu(\Lambda), \]
and $\mu(\Lambda)\geq \frac{1}{4}\mu(\Sigma)$. Now we explain why the fortification lemma of \cite{DBLP:journals/cc/DinurM18} is a special case of ours. Given a communication problem $S\subseteq (X\times Y) \times Z$, given a rectangle $A\times B \subseteq X\times Y$,  recall  $\sfL(A\times B)$ denote the protocol size of the rectangle. Now give a fixed $A\subseteq X$, set $\sfL_{A\times }(B) = \sfL(A\times B)$ where $B\subseteq Y$, similarly we can define  $\sfL_{\times B }$ for some fixed  $B\subseteq Y$. It is easy to verify that  $\sfL_{A\times }$(respectively $\sfL_{\times B}$) is a sub-additive measure over $Y$(respectively $X$). Dinur and Meir's original fortification lemma \cite{DBLP:journals/cc/DinurM18} is exactly about the two measures  $\sfL_{A\times }$ and  $\sfL_{\times B}$.

 Now consider the measure of cover number, let $\Sigma$ be any subset of $X\times Y$,  cover number $\mathsf{Cov}$ is a sub-additive measure over $\Sigma$. Note that $\Sigma$ is not necessarily a rectangle. Set $\Sigma=X\times Y$, by the  general fortification lemma,  there is a subset $\Lambda \subseteq X\times Y$ such that for any subset $\tilde{\Lambda} \subseteq\Lambda$, 
 \[ \mathsf{Cov}(\tilde{\Lambda}) \geq \frac{1}{16\log |X||Y|} \cdot  \frac{|\tilde{\Lambda}|}{|\Lambda|} \cdot  \mathsf{Cov}(S).\]
Note that when $\frac{|\tilde{\Lambda}|}{|\Lambda|} > \frac{16\log |X||Y|}{\mathsf{Cov}(S)}$, $\mathsf{Cov}(\tilde{\Lambda}) >1$, this means $\Lambda$ is a $\frac{16\log|X| |Y|}{\mathsf{Cov}(S)}$-fooling set.

Now we show how to use $\delta$--fooling set to prove direct sum  theorem. The basic idea is to construct the hardcore of the  direct sum problem from hardcore of each problem. Directly, Cartesian product of each problem's $\delta$--fooling set   is a hardcore of the direct sum of problems, but sometimes, to prove a stronger results, we need to construct a more delicate hardcore. Take the direct sum problem of cover number as an example, given two communication problems $S  \subseteq (A\times B) \times O$ and $T \subseteq (P\times Q) \times Z$, let $\Lambda$ be a $\delta$-fooling set for problem $T$, we want to lower bound the cover number of their direct sum problem $S\times T$. Given any $(p,q)  \in \Lambda$, note that $(A\times\{p\}) \times(B\times 
\{q\})$ is a rectangle of $S\times T$, and we construct the hardcore for $S\times T$ to be the collection of all such rectangles $(A\times\{p\}) \times(B\times 
\{q\})$ for every $(p,q)  \in \Lambda$. Formally, the hardcore is simply $\cup_{(p,q)\in \Lambda} (A\times\{p\}) \times(B\times 
\{q\})$.

Now we give the intuition that why this hardcore is indeed hard. At first, note that every such rectangle $(A\times\{p\}) \times(B\times 
\{q\})$ will need at least $\mathsf{Cov}(S)$ monochromatic rectangles to cover it. Since there are $|\Lambda|$ such rectangles, if we allow multiplicity, the total number to cover all such rectangles is  at least $|\Lambda|  \cdot \mathsf{Cov}(S)$.
Now we handle the problem of multiplicity since a monochromatic rectangle  $R$ could cover elements from different rectangles $(A\times\{p\}) \times(B\times 
\{q\})$ for different $(p,q) \in \Lambda$.  Fortunately, we can show  a monochromatic rectangle $R$ could cover elements from at most $\delta|\Lambda|$ different rectangles.  Denote $\{(p,q)|\exists ((a,p),(b,q)) \in R\}$ by $R|_{T}$, if $R$ is  monochromatic, then $R|_{T}$ is also monochromatic   rectangle which contains no more than $\delta|\Lambda|$ such $(p,q)$, otherwise it would  contradict that $\Lambda$ is a $\delta$--fooling set of $T$. This means a monochromatic rectangle $R$ could cover elements from at most $\delta|\Lambda|$ different rectangles, thus we need at least $|\Lambda|  \cdot \mathsf{Cov}(S)/ \delta |\Lambda| = \mathsf{Cov}(S)/ \delta$ monochromatic rectangles to cover the hardcore $\cup_{(p,q)\in \Lambda} (A\times\{p\}) \times(B\times 
\{q\})$.

\subsection{Organization of the rest of the paper}
We provide the necessary preliminaries in Section $2$. In Section $3$, we define the $\delta$-fooling set, present the general fortification lemma then apply it to the case of cover number. In Section $4$, we reprove the direct sum theorem of cover number and prove a new direct sum theorem of protocol size. Finally, in Section $5$,  we conclude and discuss some future directions.

\section{Preliminaries}
In this section, we provide some basic notations, definitions and facts. Let $\mathbb{N}$ be the set of nature number, for any $n \in \mathbb{N}$, we denote by $[n]$ the set $\{1, \ldots, n\}$. We often use bold font to indicate $\mathbf{x} \in X^n$ is a vector, and  denote the $i$-th coordinate of the vector by  $\mathbf{x}_i$. We assume the readers are familiar with the basic knowledge of communication complexity, a more detailed introduction of communication complexity can be found in textbooks such as \cite{DBLP:books/daglib/0011756,rao_yehudayoff_2020}. 
\begin{defn}[Two party communication problems]In a two party communication problem ${S} \subseteq ({X} \times {Y}) \times Z$, there are two involved players--Alice and Bob who need to solve following task: 
	Alice is given an input $x \in X$ and Bob is given an input $y\in Y$, they need to output a element $z \in Z$ such that $(x,y,z)\in S$.
\end{defn}

\paragraph{Rectangle cover and cover number}
\begin{defn}Given a communication problem ${S} \subseteq ({X} \times {Y}) \times Z$, let $R=A\times B \subset X\times Y$ be a rectangle, if for every $(x,y)\in A \times B$, $(x,y,z) \in S$, we say rectangle $R$ is monochromatic with color $z$ or $z$-monochromatic for short. Let $\Sigma \subseteq X\times Y$ and  $\chi$ be a set of monochromatic rectangles of $S$, we say $\chi$ is  a rectangle cover for $\Sigma$, or simply $\chi$ covers $\Sigma$, if for every element $(x,y) \in \Sigma$ there is a monochromatic rectangle $R\in  \chi$ such that $(x,y) \in R$. The cover number of $\Sigma$, denoted by $\mathsf{Cov}(\Sigma)$, is  the minimum number of monochromatic rectangles to cover $\Sigma$, formally, $\mathsf{Cov}(\Sigma) =\min_{\chi\text{covers }\Sigma} |\chi|.$ Particularly, when $\Sigma = X\times Y$, we simply write $\mathsf{Cov}(S)$, that is the cover number of communication problem $S$.
\end{defn}
\paragraph{Deterministic protocol}
\begin{defn}
	A deterministic protocol  $\Pi: {X} \times {Y} \rightarrow Z$ for a communication problem ${S} \subseteq \left({X} \times {Y}\right)\times Z$ is  a rooted binary tree with  following structure:
	\begin{itemize}
		\item  Every node $v$ in the tree  belongs to Alice or Bob and is associated with a rectangle $R_v =X_v \times Y_v \subseteq X\times Y$. Particularly, the root of protocol tree is associated with the  rectangle ${X} \times {Y}$.
		\item  Given an internal node $v$, let $v_0,v_1$ be two children of $v$. Recall  $v$ is associated with a rectangle $R_v =X_v \times Y_v$, if $v$ is owned by Alice, then $v_0$ is associated with $X_{v_0} \times Y_v$, $v_1$ is associated with $X_{v_1} \times Y_v$ where $X_{v_0} \cap X_{v_1} = \emptyset$ and $X_{v_0} \cup X_{v_1} =X_v$; if $v$ is owned by Bob, then $v_0$ is associated with $X_v \times Y_{v_0}$, $v_1$ is associated with $X_v \times Y_{v_1}$ where $Y_{v_0} \cap Y_{v_1} = \emptyset$ and $Y_{v_0} \cup Y_{v_1} =Y_v$.
		\item  Every leaf node $\ell$ is  associated with a  monochromatic rectangle $R_\ell$ with color $z$ and $z$ is the output of the protocol.
	\end{itemize}
\end{defn}

\paragraph{Communication complexity and protocol size}
\begin{defn}Given a protocol tree $\Pi$, its depth $\mathsf{D}(\Pi)$ is the length of the longest path from the root to a leaf in the tree. Given a communication problem $S\subseteq (X\times Y) \times Z$, the  (deterministic) communication complexity $\mathsf{C}(S)$ of communication problem $S$ is the minimum $\mathsf{D}(\Pi)$ over all protocol $\Pi$ for the problem $S$. Given a protocol tree $\Pi$, its protocol size $\mathsf{L}(\Pi)$ is the number of leaves of the tree.  Given a communication problem $S$, its complexity of protocol size  $\mathsf{L}(S)$ is minimum $\mathsf{L}(\Pi)$ over all protocol $\Pi$ for the problem $S$.
\end{defn}
It is well known that the complexity of  protocol size is sub-additive over rectangles.   Formally, we have following fact.
\begin{fact}\label{fact2.6}Given a communication problem  $S\subseteq (X\times Y) \times Z$, let $A\times B\subseteq X\times Y$ and $\sfL(A\times B)$ be the complexity of  protocol size to solve problem $S$ when restricted to rectangle $A\times B$, we have
	\begin{itemize}
		\item  $\mathsf{L}\left((A_0 \cup A_1) \times B\right) \leq  \mathsf{L}\left(A_0  \times B\right) +\mathsf{L}\left( A_1\times B\right)$. 
		\item  $\mathsf{L}\left( A\times(B_0 \cup B_1)  \right) \leq  \mathsf{L}\left(A  \times B_0\right) +\mathsf{L}\left( A\times B_1\right)$. 
	\end{itemize}
\end{fact}
\begin{fact}\label{fact2.7}\cite{DBLP:books/daglib/0011756} For  every communication problem $S$,  it  holds that
	$$
	\log \mathsf{L}(S) \leq \mathsf{C}(S) \leq 2 \log \mathsf{L}(S)
	$$
	and hence $\mathsf{C}(S)=\Theta(\log \mathsf{L}(S))$.
\end{fact}
\begin{fact}\label{fact2.8}\cite{DBLP:books/daglib/0011756}Let communication problem $F:X\times Y \rightarrow \{0,1\}$ be a function, then
	\[\log\mathsf{Cov}(F)=\Omega\left(\sqrt{\mathsf{C}(F)}\right).\]
\end{fact}
We will need  following simple but important fact which says projection of a monochromatic rectangle is still a monochromatic.
\begin{fact}\label{fact4.2}Given two communication problems $S  \subseteq (A\times B) \times O$ and $T \subseteq (P\times Q) \times Z$, let $R\subseteq (A\times P) \times (B\times Q)$ be a monochromatic rectangle of $S\times T$. Denote the set $\{(a,b)|\exists ((a,p),(b,q)) \in R\}$ by $R|_{S}$  and similarly  $\{(p,q)|\exists ((a,p),(b,q)) \in R\}$ by $R|_{T}$, we have  $R|_{S}$(respectively $R|_{T}$) is a  monochromatic rectangle of $S$(respectively $T$).
\end{fact}
\begin{proof}We prove it for $R|_S$, the case for $R|_T$ is similar. At first, we will show if  $(a,b),(a',b')$ are contained in $R|_S$, so are  $(a',b),(a,b')$.  If  $(a,b),(a',b')$ are contained in $R|_S$, there must be two elements $((a,p),(b,q	)),((a',p'),(b',q'))\in R$, that is  $((a,p),(b',q')),((a',p'),(b,q)) \in R$, which means $(a',b),(a,b')$ are contained in  $R|_S$. Furthermore,  since $R$ is monochromatic with some color $(o,z) \in O \times Z$, every $(a,b)\in R|_S$ can be colored with $o$, thus $R|_S$ is a monochromatic rectangle of $S$ as required.
\end{proof}

\section{Generalized Fooling Set and Fortification of Cover Number}
In this section, we introduce a  generalized form of  standard fooling set called the $\delta$-fooling set. 
\begin{defn}[$\delta$-fooling set]Let  $S\subseteq (X\times Y) \times Z$ be a communication problem, we call a set  $\Lambda \subseteq  X\times Y$ a $\delta$-fooling set of $S$ if for any  subset $\tilde{\Lambda} \subseteq \Lambda$ such that $
	\frac{|\tilde{\Lambda}|}{|\Lambda|} > \delta$, there is no  monochromatic rectangle that covers all elements in the subset $\tilde{\Lambda}$.
\end{defn}
The $\delta$-fooling set can be viewed as a dual form of cover number,  given a communication problem, a large cover number for this problem is equivalent to there is a $\delta$-fooling set with small $\delta$ for this problem. At first, we show the easy direction, that is  a $\delta$-fooling set with small $\delta$ implies a large cover number. This mimics the effect of  standard fool set.

\begin{prop}\label{prop3.2}Let  $S\subseteq (X\times Y) \times Z$ be a communication problem, let  $\Lambda \subseteq  X\times Y$ be a $\delta$-fooling set of $S$, then $\mathsf{Cov}(\Lambda)\geq 1/\delta$.
\end{prop}
\begin{proof}Since any monochromatic rectangle can only cover at most $\delta|\Lambda|$ elements from $\Lambda$, at least $ 1/\delta$ monochromatic rectangles are required to cover all elements from $\Lambda$.
\end{proof}

Now we present the other direction: that is a large cover number implies a $\delta$-fooling set with small $\delta$. This is achieved by applying  the general fortification lemma to the case of cover number. At first, we need a notion called sub-additive measure over set.
\begin{defn} Let $\Sigma$ be a nonempty finite set, a sub-additive measure over set  $\Sigma$ is a function $\mu: 2^\Sigma \rightarrow \mathbb{N}$ with following properties:
	\begin{itemize}
		\item semipositivity: 
		$\mu(\emptyset)=0$ and 	if $\Lambda \subseteq \Sigma$  is not empty, $\mu(\Lambda)\geq 1$,
		\item subadditivity: given two subsets $\Lambda,\Lambda^\prime \subseteq \Sigma$, 
		$\mu(\Lambda \cup \Lambda^\prime)\leq\mu(\Lambda)+\mu(\Lambda^\prime).$
	\end{itemize}
\end{defn}

Given a communication problem ${S}\subseteq ({X}\times {Y}) \times {Z}$, the cover number is a sub-additive measure over any subset $\Sigma$ of $X\times Y$. Formally we have following fact.
\begin{fact}\label{fact3.4}Give a communication problem ${S}\subseteq ({X}\times {Y}) \times {Z}$, let $\Sigma\subseteq X\times Y$  be a subset, then  cover number $\mathsf{Cov}$ according to the communication problem ${S}$ is a sub-additive measure over $\Sigma$.
\end{fact}

Next we define the notion of general fortification.
\begin{defn}Let $\Sigma$ be a nonempty finite set, $\mu$ be a sub-additive measure over set  $\Sigma$ and  $\Lambda$ be a subset of  $\Sigma$. Given any subset   $\tilde{\Lambda} \subseteq \Lambda$, we say  $\Lambda$ is $\rho$-fortified with respect to measure $\mu$, if  for any such $\tilde{\Lambda}$, it holds that
	$\mu(\tilde{\Lambda})\geq \rho \cdot \frac{|\tilde{\Lambda}|}{|{\Lambda}|}\cdot \mu({\Lambda}).$
\end{defn}

\begin{lem}[General fortification lemma]\label{lem3.6}Given a set $\Sigma$ and  a sub-additive measure $\mu$ over $\Sigma$. There exists $\Lambda \subseteq \Sigma$ such that
	\begin{itemize}
		\item $\Lambda$ is $\frac{1}{4\log |\Sigma|}$--fortified 
		\item and $\mu\left(\Lambda\right)\geq \frac{1}{4} \mu\left(\Sigma\right)$.
	\end{itemize}  
\end{lem}
The proof of the general fortification lemma is similar to its less general form in \cite{DBLP:journals/cc/DinurM18} and is deferred to  Appendix \ref{apd1}. Now we can apply it to Fact \ref{fact3.4} to  obtain  the fortification of cover number.
\begin{prop}[Fortification of cover number]\label{prop3.7}Let  $S\subseteq (X\times Y) \times Z$ be a communication problem,  there exists $\Lambda \subseteq  X\times Y$ such that for any subset $\tilde{\Lambda} \subseteq \Lambda$, we have
	$$\mathsf{Cov}(\tilde{\Lambda}) \geq
	\frac{|\tilde{\Lambda}|}{|\Lambda|}
	\cdot\frac{ \mathsf{Cov}(S)}{16\log|X| |Y|}.$$
Particularly, when $	\frac{|\tilde{\Lambda}|}{|\Lambda|}> \frac{16\log|X||Y|}{\mathsf{Cov}(S)}$, we have  $\mathsf{Cov}(\tilde{\Lambda}) > 1$, this means $\Lambda$ is a $\frac{16\log|X| |Y|}{\mathsf{ Cov}(S)}$-fooling set.
\end{prop}

\section{Direct Sum Theorems from Fortification}
\subsection{Direct sum theorem of cover number, revisit}
In this section, we revisited the direct sum problem of cover number and present an alternative proof.\footnote{An anonymous reviewer points out that \cite{DBLP:books/daglib/0011756} give a proof of this theorem with a similar fashion but in a different way. Nevertheless, we keep our proof here as another alternative.} In our proof, we only use  Proposition \ref{prop3.7} and a double counting argument. Formally, we have following theorem.
\begin{thm}\label{thm4.1}	Given two communication problems $S  \subseteq (A\times B) \times O$ and $T \subseteq (P\times Q) \times Z$, let $\Lambda$ be a $\delta$-fooling set for problem $T$, we have 
	\begin{align*}
	\mathsf{Cov}\left(S\times T\right)
	\geq \mathsf{Cov}\left(S\right) /
	\delta.
	\end{align*}
	Particularly, by Proposition \ref{prop3.7}, we have
	\begin{align*}
	\log \mathsf{Cov}\left(S\times T\right)
	\geq  \log \mathsf{Cov}\left(S\right) 
	+ \log\mathsf{Cov}(T)
	-\log\log|P||Q|-4.
	\end{align*}
\end{thm}

\begin{proof}(of Theorem \ref{thm4.1})Let $\chi$ be a rectangle cover of $S\times T$, we will prove $|\chi| \geq \mathsf{Cov}\left(S\right) /
	\delta $.  Given any $(p,q)  \in \Lambda$, $(A\times\{p\}) \times(B\times 
\{q\})$ is a rectangle of $S\times T$, every such rectangle will need at least $\mathsf{Cov}(S)$ monochromatic rectangle to cover it. Since there are $|\Lambda|$ such rectangles, if we allow multiplicity, the total number to cover all such rectangles is  at least $|\Lambda|  \cdot \mathsf{Cov}(S)$. Formally, let  $R\in \chi$, denote \[\left(R  \cup (A\times\{p\}) \times(B\times 
\{q\}) \right)|_T\] by $R_{(p,q)}$, we have
\[\sum_{R\in \chi} \sum_{(p,q)\in \Lambda} \mathbf{1}_{R_{(p,q)} \neq \emptyset}   =   \sum_{(p,q)\in \Lambda} \left( \sum_{R\in \chi} \mathbf{1}_{R_{(p,q)} \neq \emptyset} \right) \geq  |\Lambda|  \cdot\mathsf{Cov}(S) \]

Now we handle the problem of multiplicity since a monochromatic rectangle  $R$ in $\chi$ could cover elements from different rectangles $(A\times\{p\}) \times(B\times 
\{q\})$ for different $(p,q) \in \Lambda$, but fortunately  a monochromatic rectangle $R$  in $\chi$ could cover elements from at most $\delta|\Lambda|$ different rectangles. Formally, we have 
\[ \sum_{(p,q)\in \Lambda} \mathbf{1}_{R_{(p,q)} \neq \emptyset} \leq  \delta|\Lambda|, \]
if not, by Fact \ref{fact4.2}, we know that $R|_T$ is a monochromatic rectangle which contains more than $\delta|\Lambda|$ such $(p,q)$, this contradicts that $\Lambda$ is a $\delta$-fooling set of $T$. Finally, we have 
\[|\chi| = \sum_{R\in \chi} \mathbf{1} \geq    \sum_{R\in \chi} \frac{\sum_{(p,q)\in \Lambda} \mathbf{1}_{R_{(p,q)} \neq \emptyset} }{ \delta|\Lambda|}  \geq  |\Lambda| \cdot \mathsf{Cov}(S) /\delta|\Lambda| = \mathsf{Cov}(S) / \delta . \]
\end{proof}
\subsection{A direct sum  theorem of protocol size}
In this section, we prove a new direct sum theorem  about complexity of protocol size. The proof is inspired by  ideas in   \cite{DBLP:journals/cc/DinurM18}. Formally, we have following theorem.

\begin{thm}\label{thm4.3}	Given two communication problems $S  \subseteq (A\times B) \times O$ and 
	$T \subseteq (P\times Q) \times Z$,  let $\Lambda$ be a $\delta$-fooling set  of $T$, we have 
	\begin{align*}
	\mathsf{L}\left(S\times T\right)
	=  \mathsf{L}\left(S\right) /
	\delta.
	\end{align*}
	Particularly, by Proposition \ref{prop3.7},  we have 
	\begin{align*}
	\log \mathsf{L}\left(S\times T\right)
	=  \log \mathsf{L}\left(S\right) 
	+ \log\mathsf{Cov}(T)
	-\log\log|P||Q|-4.
	\end{align*}
\end{thm}

Let's  recall some  notations. Let $R\subseteq (A\times P) \times (B\times Q)$ be a  rectangle of $S\times T$. Denote the set $\{(a,b)|\exists ((a,p),(b,q)) \in R\}$ by $R|_{S}$  and similarly  $\{(p,q)|\exists ((a,p),(b,q)) \in R\}$ by $R|_{T}$. Furthermore, given any $(p,q)  \in \Lambda$, denote $\left(R  \cup (A\times\{p\}) \times(B\times 
\{q\}) \right)|_S$ by $R_{(p,q)}$.

\paragraph{The sub-additive measure over protocol tree.}  We introduce the definition of a sub-additive measure over protocol tree as follows.
\begin{defn}[\cite{DBLP:journals/cc/DinurM18}]Given a rooted binary tree $T$ and let $V$ be the set of nodes of tree $T$, we say that $\phi: V \rightarrow \mathbb{N}$ is a sub-additive measure on $T$ if for every vertex $v$ with children $v_0$ and $v_1$ in $T$ it holds that $\phi(v) \leq  \phi(v_0) +\phi(v_1)$.
\end{defn}
Now we define a special sub-additive measure over protocol tree following the similar idea in  \cite{DBLP:journals/cc/DinurM18}.
\begin{defn}Given a protocol tree $\Pi$ for $S \times T$,  let $\pi$ be a node in the protocol tree $\Pi$, denote the rectangle associated with the node $\pi$ by $R_\pi \subseteq (A\times P) \times (B\times Q)$, let $\Lambda$ be a $\delta$-fooling set for cover number of $T$,  we define a measure $\phi$ on all such $\pi$ as follows: 
	\[
	\phi(\pi
	)=\frac{1}{|\Lambda|} \sum_{
		( p,q)\in \Lambda}
	\mathsf{L}\left( {R_\pi}_{(p,q)}\right).
	\]
\end{defn}
Recall that ${R_\pi}_{(p,q)} =\left({R_\pi}  \cup (A\times\{p\}) \times(B\times 
\{q\}) \right)|_T$.	Intuitively,  the measure $\phi$ is just the average complexity of all  such rectangles  ${R_\pi}_{(p,q)}$ where $(p, q) \in \Lambda$. It easy to verify following fact about the measure $\phi$.
\begin{fact}\label{fact4.6}The  measure $\phi$ is a sub-additive measure on protocol tree  $\Pi$. Furthermore, 
	$\phi$ assigns $\mathsf{L}(A\times B)=\mathsf{L}(S)$ to the root of $\Pi$.
\end{fact}
\begin{proof} At first, the measure $\phi$
	is sub-additive since, by Fact \ref{fact2.6}, for every fixed $(p,q)$, the measure $\mathsf{L}\left({R_{\pi}}_{(p,q)}\right)$ is sub-additive over the protocol tree $\Pi$. Furthermore, $\phi$ assigns $\mathsf{L}({A} \times {B})$ to the root of $\Pi$, since when $\pi$ is the root, ${R_{\pi}}_{(p,q)}$ simply is ${A} \times {B}$, for every $(p,q)$.
\end{proof}
We will also need following fact which claims for each leaf in the protocol tree, its measurement is small.
\begin{fact}\label{fact4.7}Given a protocol $\Pi$ which solves ${S}\times {T}$ and $\ell$ is a leaf of $\Pi$, then $\phi(\ell) \leq  \delta$.
\end{fact}
\begin{proof}Let $R_\ell$ be the rectangle associated with the leaf $\ell$. Since $\ell$ is a leaf, $R_\ell$ is monochromatic and $\mathsf{L}(R_\ell) \leq 1$, this means for every $(p,q)\in  \Lambda$, $0\leq \mathsf{L}({R_\ell}_{(p,q)}) \leq 1$,  and  $R_\ell|_S,R_\ell|_T$ are also monochromatic. Let $\tilde{\Lambda}$ be the set of all $(p,q)$ such that $\mathsf{L}({R_\ell}_{,(p,q)}) \neq 0$. Since  $\tilde{\Lambda}$ is contained in monochromatic rectangle $R|_T$, this means $|\tilde{\Lambda}| \leq \delta |\Lambda|$ due to $\Lambda$ is  a $\delta$-fooling set  of $T$. Now we are ready to bound $\phi(\ell)$, that is 
	\begin{align*}
		\phi(\ell
	)=\frac{1}{|\Lambda|} \sum_{
		( p,q)\in \Lambda}
	\mathsf{L}({R_\ell}_{(p,q)})
	= \frac{1}{|\Lambda|} \sum_{
		( p,q)\in \Lambda}
	\mathbf{1}_{\mathsf{L}({R_\ell}_{(p,q)})\neq 0}
	= \frac{\tilde{|\Lambda|}}{|\Lambda|}\leq \delta.
	\end{align*}
\end{proof}
Now we are ready to prove our theorem about protocol size.
\begin{proof}(of Theorem \ref{thm4.3})Given a  protocol $\Pi$ for $S \times T$, let $r$ be the root of protocol $\Pi$, by the subadditivity of $\phi$,
	\[\phi(r) \leq \sum_{\ell \text{ is a leaf}} \phi(\ell).
	\]
	By Fact \ref{fact4.6} and  Fact \ref{fact4.7}, we have
		$\sfL(S) \leq \sfL(\Pi)\cdot\delta,
	$
	that is 
	$\sfL(\Pi) \geq  \sfL(S)/\delta.
	$
\end{proof}

By Theorem \ref{thm4.3}, Fact \ref{fact2.7} and Fact \ref{fact2.8}, we have following corollary.
\begin{cor}Given a communication problem $F:A \times B \rightarrow 
	\{0,1\}$, 
	\begin{align*}
	\log\mathsf{L}\left(F\times F\right)\geq
	\log \mathsf{L}\left(F\right) +\Omega\left(\sqrt{\log\mathsf{L}\left(F\right)}\right)-\log\log|A||B| -4.
	\end{align*}	
\end{cor}

\section{Conclusion and Discussion}
We conclude with some discussion about our results and future direction. The first question is can we fortify other measures in communication complexity besides cover number? It was pointed out by \cite{DBLP:journals/cc/DinurM18} that it is impossible to fortify both sides of the rectangle simultaneously. So we should consider other measures which avoid such large gap. An interesting question is whether  we  can fortify any useful measures in randomized communication complexity? Note that if we relax rectangle cover to cover of  nearly monochromatic rectangles, the lemma also works, the issue here is we don't know whether a small cover number of  nearly monochromatic rectangles implies a small randomize complexity. Besides measures in  communication complexity, we can apply the fortification lemma to  other measures such as the measure of entropy  $\mathsf{H}$. Let $\mathbf{X}_1, \mathbf{X}_2,\cdots, \mathbf{X}_n$ be $n$ joint distributed random variables, the measure  entropy  $\mathsf{H}$ is sub-additive over these random variables, thus can be fortified. It is interesting that whether these fortifications lead to further applications.

\section*{Acknowledgment}
The author wants to thank the anonymous reviewer for pointing out a gap in  an early version of this paper and  other helpful comments.

\bibliographystyle{alpha}
\bibliography{hw}
\appendix
\section{The Proof of General Fortification Lemma}\label{apd1}
The general fortification lemma is proved in a similar way  to its less general form in \cite{DBLP:journals/cc/DinurM18}. At first,  we  show a  so called the weak fortification.

\begin{prop}\label{propa.1}Given a set $\Sigma$, a sub-additive measure $\mu$ over  $\Sigma$ and $0<\rho<1$, there exists $\Lambda_1 \subseteq \Sigma$ such that:
	\begin{itemize}
		\item for every $\tilde{\Lambda} \subseteq \Lambda_1$, it holds that $\mu(\tilde{\Lambda}) \geq \rho \cdot \frac{|\tilde{\Lambda}|}{|\Sigma|}\cdot \mu(\Sigma).$
		\item $\mu\left(\Lambda_1\right) \geq(1-\rho) \cdot \mu(\Sigma)$.
	\end{itemize}
\end{prop}
\begin{proof}Let $\Lambda_{\max} \subseteq \Sigma $ 	be a maximal subset under the order of set inclusion that satisfies
	\begin{equation}\label{eqa.1}
			\mu\left(\Lambda_{\max}\right)<\rho \cdot \frac{\left|\Lambda_{\max}\right|}{|\Sigma|} \cdot \mu(\Sigma).
	\end{equation}
	Let $\Lambda_1  \stackrel{\text{def }}{=} \Sigma-\Lambda _{\text {max}}$, by the subadditivity of measure $\mu$, we have
	\[\mu(\Sigma)=\mu(\Lambda_1 \cup \Lambda_{\max})
	\leq\mu(\Lambda_1)+\mu(\Lambda_{\max}),\]
	thus by rearranging above inequality, we have obtained the second item in this proposition, that is
	\[\mu\left(\Lambda_1\right) \geq
	\mu(\Sigma)-\mu(\Lambda_{\max})>	\mu(\Sigma)- \rho \cdot \frac{\left|\Lambda_{\max}\right|}{|\Sigma|} \cdot \mu(\Sigma)
	\geq 
	(1-\rho) \cdot \mu(\Sigma).\]
	Now to obtain the first item of this proposition, for the sake of contradiction, suppose that there is a nonempty subset $\tilde{\Lambda}\subseteq \Lambda_1$, such that $\mu(\tilde{\Lambda}) < \rho \cdot \frac{|\tilde{\Lambda}|}{|\Sigma|} \cdot \mu(\Sigma)$. Then, this would imply that
	$$
	\begin{aligned}
	\mu\left(\tilde{\Lambda} \cup \Lambda_{\max }\right)  & \leq \mu(\tilde{\Lambda})+\mu\left(\Lambda_{\max } \right),\text{by subadditivity of $\mu$} \\
	&<\rho \cdot \frac{|\tilde{\Lambda}|}{|\Sigma|} \cdot \mu(\Sigma )+\rho \cdot \frac{\left|\Lambda_{\max }\right|}{|\Sigma|} \cdot \mu(\Sigma ),\text{ by  assumptions on $\tilde{\Lambda}$ and $\Lambda_{\max }$} \\
	&=\rho \cdot \frac{\left|\tilde{\Lambda} \cup \Lambda_{\max }\right|}{|\Sigma|} \cdot \mu(\Sigma ).
	\end{aligned}
	$$
	It turns out that $\tilde{\Lambda} \cup \Lambda_{\max }$ is a set that satisfies Inequality (\ref{eqa.1}) and that strictly contains $\Lambda_{\max },$ thus contradicting the maximality of $\Lambda_{\max }$. Hence, no such set $\tilde{\Lambda}$ exists, the first item of this Proposition holds.
\end{proof}

The above proposition is weak because the measure of $\tilde{\Lambda}$ is propositional to its density in $\Sigma$ rather than $\Lambda$. To proceed, we need following fact about ``inverse fortification''.
\begin{prop}\label{propa.2}Given a set $\Sigma$ and a sub-additive measure $\mu$ over $\Sigma$, for every $c \geq 1,$ there exists a subset $\Lambda_0 \subseteq \Sigma$ such that for every $\tilde{\Lambda} \subseteq \Lambda_0$ it holds that
	\begin{equation} \label{eqa.2}
		\frac{|\tilde{\Lambda}|}{\left|\Lambda_0\right|} \geq\left(\frac{\mu(\tilde{\Lambda} )}{\mu\left(\Lambda_0 \right)}\right)^{c}
	\end{equation}
	and 
	\begin{equation}\label{eqa.3}
	\mu\left(\Lambda_0 \right) \geq \left(\frac{1}{|\Sigma|}\right)^{\frac{1}{c}} \cdot \mu(\Sigma).
	\end{equation}
\end{prop}
\begin{proof}At first, we set $\Lambda_0$ to be a minimal set under the order of set inclusion that satisfies
	$$
	\frac{\left|\Lambda_0\right|}{|\Sigma|} \leq\left(\frac{\mu\left(\Lambda_0 \right)}{\mu(\Sigma )}\right)^{c}
	$$
	Observe that $\Lambda_0$ indeed satisfies Inequality (\ref{eqa.2}):  if not, there must be a proper subset $\tilde{\Lambda} \subsetneq \Lambda_0$ which satisfies
	$$
	\frac{|\tilde{\Lambda}|}{\left|\Lambda_0\right|}<\left(\frac{\mu(\tilde{\Lambda} )}{\mu\left(\Lambda_0 \right)}\right)^{c}.
	$$
	and this would have implied that
	$$
	\begin{aligned}
	\frac{|\tilde{\Lambda}|}{|\Sigma|} &=\frac{|\tilde{\Lambda}|}{\left|\Lambda_0\right|} \cdot \frac{\left|\Lambda_0\right|}{|\Sigma|} 
	<\left(\frac{\mu(\tilde{\Lambda} )}{\mu\left(\Lambda_0 \right)}\right)^{c} \cdot\left(\frac{\mu\left(\Lambda_0 \right)}{\mu(\Sigma )}\right)^{c}
	=\left(\frac{\mu(\tilde{\Lambda} )}{\mu(\Sigma )}\right)^{c}
	\end{aligned}
	$$
	thus contradicting the minimality of $\Lambda_0$. Now it remains to show that $\Lambda_0$ satisfies Inequality (\ref{eqa.3}). Recall that we set $\Lambda_0$ to satisfy
	$$
	\frac{\left|\Lambda_0\right|}{|\Sigma|} \leq\left(\frac{\mu\left(\Lambda_0 \right)}{\mu(\Sigma )}\right)^{c},
	$$
	by rearranging above inequality, we have
	$$
	\begin{aligned}
	\mu\left(\Lambda_0 \right)  
	\geq\left(\frac{\left|\Lambda_0\right|}{|\Sigma|}\right)^{\frac{1}{c}} \cdot \mu(\Sigma ) 
	\geq\left(\frac{1}{|\Sigma|}\right)^{\frac{1}{c}} \cdot \mu(\Sigma ) 
	\end{aligned}
	$$
	as required.
\end{proof}

Finally, we are ready to prove our general fortification lemma.
\begin{proof}(Proof of Lemma \ref{lem3.6}). 
Our goal is to find a subset $\Lambda \subseteq \Sigma$ such that 
\begin{itemize}
	\item $\Lambda $ is $\frac{1}{4\log|\Sigma|}$-fortified,
	\item and  $\mu\left(\Lambda \right) \geq \frac{1}{4} \cdot \mu(\Sigma )$
\end{itemize} 
Now  we  apply Proposition \ref{propa.2} to $\Sigma$ with $c= \log|\Sigma|$ and obtain a subset $\Lambda_0 \subseteq$ $\Sigma .$ Then, we apply Proposition \ref{propa.1} to $\Lambda_0$ with $\rho=\frac{1}{2\log|\Sigma|},$ thus obtaining a subset $\Lambda_1 \subseteq \Lambda_0$. Finally, we choose $\Lambda$ to be $\Lambda_1$.
We prove that $\Lambda$ has the required properties. At first,  we show  $\mu\left(\Lambda \right) \geq \frac{1}{4} \cdot \mu(\Sigma )$. Note that by Proposition \ref{propa.1}, it holds that
	\begin{equation}\label{eqa.4}
	\mu\left(\Lambda \right) \geq\left(1-\frac{1}{2\log|\Sigma|}\right) \cdot \mu\left(\Lambda_0 \right) \geq \frac{1}{2} \cdot \mu\left(\Lambda_0 \right)
	\end{equation}
	and that by Proposition \ref{propa.2}, it holds that
	$$
	\mu\left(\Lambda_0 \right) \geq\left(\frac{1}{|\Sigma|}\right)^{\frac{1}{\log\Sigma}} \cdot \mu(\Sigma )  \geq \frac{1}{2} \cdot \mu(\Sigma).
	$$
	Therefore,
	$$
	\mu\left(\Lambda \right) \geq \frac{1}{4} \cdot \mu(\Sigma )
	$$
	as required. It remains to show $\Lambda$ is $\frac{1}{4\log|\Sigma|}$--fortified. Let $\tilde{\Lambda} \subseteq \Lambda .$ By Proposition \ref{propa.1}, it holds that
$$
\mu(\tilde{\Lambda} ) \geq \frac{1}{2\log|\Sigma|} \cdot \frac{|\tilde{\Lambda}|}{\left|\Lambda_0\right|} \cdot \mu\left(\Lambda_0 \right) \geq \frac{1}{2\log|\Sigma|} \cdot \frac{\left|\Lambda\right|}{\left|\Lambda_0\right|} \cdot \frac{|\tilde{\Lambda}|}{\left|\Lambda\right|} \cdot \mu\left(\Lambda \right)
$$
Next, by Proposition \ref{propa.2}, it holds that
$$
\begin{aligned}
\frac{\left|\Lambda\right|}{\left|\Lambda_0\right|} \geq\left(\frac{\mu\left(\Lambda \right)}{\mu\left(\Lambda_0 \right)}\right)^{\log|\Sigma|} &\geq\left(1-\frac{1}{2 \log|\Sigma|}\right)^{\log|\Sigma|} \text{, by Equation \ref{eqa.4}}\\
&\geq \frac{1}{2}.
\end{aligned}
$$
Thus,
$$
\mu(\tilde{\Lambda} ) \geq \frac{1}{4\log|\Sigma|} \cdot \frac{|\tilde{\Lambda}|}{\left|\Lambda\right|} \cdot \mu\left(\Lambda \right).
$$
This means $\Lambda$ is $\frac{1}{4\log|\Sigma|}$--fortified as required.
\end{proof}

\end{document}